\documentstyle[12pt]{article}

\newcommand{\eq}[1]{eq.\,(\ref{eq#1})}

\title{General equation for Zeno-like effects in spontaneous
exponential decay}
\author{Alexander D.\ Panov\\
{\small Russian Research Center ``Kurchatov's Institute''}\\
{\small 123182 Kurchatov Sq., Moscow, Russia}\\
{\small e-mail: \tt A.Panov@relcom.ru}}
\date{}

\begin{document}

\maketitle

\begin{abstract}
It was shown that different mechanisms of perturbation of spontaneous
decay constant: inelastic interaction of emitted particles with
particle detector, decay onto an unstable level, Rabi transition from
the final state of decay (electromagnetic field domination) and some
others are really the special kinds of one general effect ---
perturbation of decay constant by dissipation of the final state of
decay. Such phenomena are considered to be Zeno-like effects and
general formula for perturbed decay constant is deduced.
\end{abstract}

\begin{center}
\end{center}

\section{Introduction}
\label{INTRODUCTION}

The term ``quantum Zeno paradox'' was introduced in
\cite{ZEN_SUDARSHAN77A,ZEN_SUDARSHAN77B}. It was argued there that an
unstable particle which is continuously observed to see whether it
decays will never be found to decay. Analogous ideas were also
discussed in some earlier works \cite{ZEN_BESKOW67,ZEN_GHIRARDI74}
(for review see \cite{ZEN_HOME97}).  Continuous observations (or
measurements) of a system were described phenomenologically in
\cite{ZEN_SUDARSHAN77A,ZEN_SUDARSHAN77B,ZEN_BESKOW67,ZEN_GHIRARDI74},
as a sequence of very frequent instantaneous collapses of system's
wave function. Later the idea of quantum Zeno paradox had been
developed in two main directions.  First, the idea was applied to
forced transitions of Rabi type between discrete levels
\cite{ZEN_COOK88} and was experimentally proved in this form
\cite{ZEN_ITANO90}.  As a result quantum Zeno paradox is considered
now as a real event (quantum Zeno effect, QZE), not as a paradox.
Second, it was recognized that phenomenological description of QZE
(using projection postulate and wave function collapses) is not
necessary.  Dynamical considerations of QZE were presented
\cite{ZEN_HOME97,ZEN_KRAUS81,ZEN_SUDBERY84,ZEN_JOOS84,ZEN_CHUMAKOV95,%
ZEN_PASCAZIO94,PANOV96F,ZEN_PASCAZIO97,ZEN_MENSKY99A} and it was
shown that the main features of QZE found earlier may be reproduced.
It was also argued that dynamical consideration of QZE could show
some essentially new features of this phenomenon \cite{PANOV96F}.

The situation of continuous observation of spontaneous exponential
decay is especially interesting. It was argued that it is impossible
to observe QZE in spontaneous decay
\cite{ZEN_GHIRARDI79,ZEN_KRAUS81,ZEN_COOK88}. But using a dynamical
approach to QZE, it was speculated
\cite{PANOV96F,ZEN_PASCAZIO97,ZEN_MENSKY99A} that perturbations of
decay rate could take place in principle and more over they could be
strong \cite{PANOV96F}. In each of three papers
\cite{PANOV96F,ZEN_PASCAZIO97,ZEN_MENSKY99A} mechanisms of
perturbation of decay rate seem to be quite different.  In
\cite{PANOV96F} the inelastic interaction of emitted particles with a
particle detector was analyzed.  In \cite{ZEN_PASCAZIO97} the forced
electromagnetic transition from the final discrete state of decay to
another (third) discrete state was studied. And in
\cite{ZEN_MENSKY99A} the spontaneous decay of the final discrete
state to another discrete state was considered. Three different
formulae describing QZE were deduced in
\cite{PANOV96F,ZEN_PASCAZIO97,ZEN_MENSKY99A} for these three
different mechanisms. But there are common points in all these
processes.  Measurement mechanism has no direct influence on the
initial state of the system (undecayed particle) in all cases.  But
as soon as the system comes to the final state of decay, it starts to
interact with some outer systems (devices or fields) that rapidly
leads to destruction of the final state of decay. The final state
of decay means the united state of decaying system with outer
systems.  Consequently, both changing of the outer systems states
and changing of the decaying system states after decay is a
destruction of the final state.  In the present paper we show that
all three mechanisms of perturbation of decay rate described in
\cite{PANOV96F,ZEN_PASCAZIO97,ZEN_MENSKY99A} are special kinds of a
general mechanism, connected with destruction of the final decay
state, and we deduce the general formula for it.  Then the formulae
of \cite{PANOV96F,ZEN_PASCAZIO97,ZEN_MENSKY99A} turn out to be the
special cases of our new general formula.

\section{General formula for Zeno-like effects}
\label{EVALUATION}

Let's consider some quantum system $S$, being in the initial state
$|\Psi_0\rangle$ at the time $t = 0$. Undisturbed Hamiltonian of the
system $S$ is $H_0$, $H_0|\Psi_0\rangle = {\cal E}_0|\Psi_0\rangle$
where ${\cal E}_0$ is the initial eigenenergy.  Under influence of
perturbation $V$, the system transits from the initial discrete state
$|\Psi_0\rangle$ to continuum of states $\{|\xi\rangle\}$ which is
orthogonal to $|\Psi_0\rangle$.  We consider that
\begin{displaymath}
   V = \sum_\xi \left( |\xi\rangle\langle\Psi_0|v(\xi)
       +|\Psi_0\rangle\langle\xi|v^*(\xi) \right),
   \label{eq1}
\end{displaymath}
where $v(\xi) = \langle\xi|V|\Psi_0\rangle$ is the matrix element for
the transition. We can suppose without restriction of generality that
$\langle\Psi_0|V|\Psi_0\rangle = 0$. We suppose also that the
perturbation $V$ is time-independent and small, thus the transition
$|\Psi_0\rangle\to\{|\xi\rangle\}$ may be approximated by a
spontaneous exponential decay for sufficiently large times.

Let's suppose that in addition to the small perturbation $V$ there
exist another interaction Hamiltonian $W(t)$, which is not small and
has dependence on time. Thus, the full Hamiltonian of the system $S$
is $H = H_0 + V + W(t)$. The interaction $W(t)$ has the feature
\begin{displaymath}
   W(t)|\Psi_0\rangle = 0,
   \label{eq2}
\end{displaymath}
i.~e.\ it does not influence the initial state of the system. But
$W(t)$ may cause a transition from subspace $\{|\xi\rangle\}$ to
another subspace $\{|\eta\rangle\}$ which is orthogonal to both
$|\Psi_0\rangle$ and $\{|\xi\rangle\}$. Let $\Gamma$ be the decay
constant of the state $|\Psi_0\rangle$. So, what is perturbed value
of $\Gamma$ if we take into account the interaction $W(t)$?

We find the no-decay amplitude
\begin{equation}
   F(t) = \langle \Psi_0|\Psi(t)\rangle e^{i{\cal E}_0 t}
   \quad\quad (\hbar = 1).
   \label{eq3}
\end{equation}
To find it we solve the Shr\"odinger equation
\begin{equation}
   |\dot \Psi(t)\rangle = -i(H_0 + V + W(t))|\Psi(t)\rangle;\quad
   |\Psi(0)\rangle = |\Psi_0\rangle.
   \label{eq4}
\end{equation}
We cannot apply the perturbation theory for $W(t)$, because this
interaction is not small, but we can do this with respect to $V$.
First let's consider the Shr\"odinger equation for the Hamiltonian
without interaction $V$:
\begin{equation}
   |\dot \Psi(t)\rangle = -i(H_0 + W(t))|\Psi(t)\rangle\quad
   \label{eq5}
\end{equation}
and let the solution of \eq{5} be
\begin{displaymath}
   |\Psi(t)\rangle = U(t,t_1)|\Psi(t_1)\rangle.
   \label{eq6}
\end{displaymath}
Let's introduce the interaction picture as
\begin{displaymath}
   |\Psi_I(t)\rangle = U^+(t,0)|\Psi(t)\rangle.
   \label{eq7}
\end{displaymath}
Then \eq{4} may be rewritten as
\begin{equation}
   |\dot\Psi_I(t)\rangle = -iV_I(t)|\Psi_I(t)\rangle, \quad
   |\Psi_I(0)\rangle = |\Psi_0\rangle
   \label{eq8}
\end{equation}
where $V_I(t)$ is the potential $V$ in the interaction picture:
\begin{displaymath}
   V_I(t) = U^+(t,0) V U(t,0).
   \label{eq9}
\end{displaymath}
In the second order of perturbation theory we easily find an
equation for derivation of no-decay amplitude from \eq{8}:
\begin{equation}
   \frac{dF}{dt} =
   -\int_0^t dt_1 \langle\Psi_0|V_I(t)V_I(t_1)|\Psi_0\rangle.
   \label{eq10}
\end{equation}
Let $\omega_\xi$ be the eigenenergy of the state $|\xi\rangle$:
$H_0|\xi\rangle = \omega_\xi|\xi\rangle$. It is not difficult to
show that the matrix element under the integral in \eq{10} may be
rewritten as
\begin{equation}
  \langle\Psi_0|V_I(t)V_I(t_1)|\Psi_0\rangle =
  \sum_\xi e^{i({\cal E}_0 - \omega_\xi)(t-t_1)} |v(\xi)|^2 D(t,t_1),
  \label{eq11}
\end{equation}
where we have introduced the {\em dissipation function}:
\begin{equation}
   D(t,t_1) =
   \frac{\langle\Psi_0|V U(t,t_1) V|\Psi_0\rangle}
   {\langle\Psi_0|V \exp[-iH_0(t-t_1)] V|\Psi_0\rangle}
   \label{eq12}
\end{equation}
The dissipation function describes the dissipation of final decay
states caused by the interaction $W(t)$. It is easily to see that if
$W(t) = 0$ then $D(t,t_1) \equiv 1$. Let the index $\xi$ of the state
be the set of the eigenenergy $\omega$ of the state and some other
quantum numbers $\alpha$: $|\xi\rangle = |\omega, \alpha\rangle$.
Let's introduce the function $M(\omega)$ as follows:
\begin{displaymath}
   M(\omega) = \sum_\alpha |v(\omega,\alpha)|^2
   \label{eq13}
\end{displaymath}
and then change $\sum_\omega$ to $\int d\omega$. Then \eq{11} is
\begin{equation}
  \langle\Psi_0|V_I(t)V_I(t_1)|\Psi_0\rangle =
  \int d\omega M(\omega) e^{i({\cal E}_0 - \omega)(t-t_1)} D(t,t_1).
  \label{eq14}
\end{equation}
For sufficiently large (but not very large) times $F(t) =
\exp(-\gamma t) \simeq 1 - \gamma t$, consequently $\gamma = -dF/dt$.
To obtain the decay constant of the state $|\Psi_0\rangle$: $\Gamma =
2{\rm Re}\gamma$ we substitute \eq{14} in \eq{10} and formally tend
$t$ to infinity supposing that this limit exists. We deduce
\begin{equation}
  \Gamma = 2\pi\int d\omega M(\omega) \Delta(\omega - {\cal E}_0)
  \label{eq15}
\end{equation}
where the function $\Delta(\epsilon)$ is defined as
\begin{equation}
  \Delta(\epsilon) = \frac{1}{\pi}{\rm Re}
  \lim_{t\to\infty}\int_0^t dt_1 e^{-i\epsilon(t-t_1)}D(t,t_1).
  \label{eq16}
\end{equation}
If $W(t) = 0$, it is easily to see that $\Delta(\epsilon) =
\delta(\epsilon)$, where $\delta(\epsilon)$ is usual Dirac's
delta-function. Then \eq{15} is transformed to $\Gamma_0 = 2\pi
M({\cal E}_0)$, i.~e.\ to usual Fermi's Golden Rule, as one could
expect.  Thus, \eq{15} is a generalization of usual Golden Rule for
case of unstable final states of decay.  The main difference between
usual Golden Rule and \eq{15} is that in the first case $\Gamma_0$ is
expressed through the single value of function $M(\omega)$, but in
the second case  $\Gamma$ is expressed through convolution of
$M(\omega)$ with spreaded function $\Delta(\omega-{\cal E}_0)$.  Now
we use the \eq{15} for studying of some particular systems.

\section{Detection of emitted particles}
\label{SCATTERING}

Let's suppose that some system $X$ (for example, an atom) transits
spontaneously from the initial exited state $|x_0\rangle$ to ground
state $|x_1\rangle$ emitting some particle $Y$ (a photon or an
electron). We consider this particle as a separate quantum system,
which is initially in the ground (vacuum) state $|y_0\rangle$ and
then transits to continuum $|\omega^Y,\alpha^Y\rangle$, where
$\omega^Y$ is energy of state and $\alpha^Y$ represents all other
quantum numbers.  Particle $Y$ inelastically scatters on a third
system $Z$  (a surrounding media or a particle detector) due to
time-independent interaction $W$.  As a result, system $Z$ transits
from the initial ground state $|z_0\rangle$ to continuum
$|\zeta\rangle$, and this process may be considered as registration
of decay. We suppose that interaction $V$ does not act on system $Z$
and that interaction $W$ does not act on system $X$.  It is a special
case of situation described in Section~\ref{EVALUATION} and we can
write:
\begin{eqnarray}
   S & = & X \otimes Y \otimes Z
   \label{eq17}\\
   H_0 & = &
   H^X_0 \otimes I_Y \otimes I_Z +
   I_X \otimes H^Y_0 \otimes I_Z +
   I_X \otimes I_Y \otimes H^Z_0
   \label{eq18}\\
   |\Psi_0\rangle & = &
   |x_0\rangle \otimes |y_0\rangle \otimes |z_0\rangle
   \equiv |x_0 y_0 z_0 \rangle
   \label{eq19}\\
   {\cal E}_0 & = & \omega^X_0 + \omega^Y_0 + \omega ^Z_0
   \label{eq20}\\
   V & = & V_{XY} \otimes I_Z; \quad
   W  =  I_X \otimes W_{YZ}
   \label{eq21}\nonumber\\
   \{|\xi\rangle\} & =  &
   \{|x_1\rangle|\omega^Y,\alpha^Y\rangle|z_0\rangle\}; \quad
   \{|\eta\rangle\} =
   \{|x_1\rangle|\omega^Y,\alpha^Y\rangle|\zeta\rangle\}
   \label{eq22}\nonumber\\
   \omega & = & \omega^X_1 + \omega^Y + \omega^Z_0
   \label{eq23}\\
   U(t,t_1) & = & \exp[-i(H_0 + W)(t-t_1)]
   \label{eq23p}
\end{eqnarray}
where notations are obvious. Using the notations
\begin{eqnarray*}
   H^{YZ}_0 = H^Y_0 \otimes I_Z + I_Y \otimes H^Z_0,
   \\
   |\tilde y\rangle =
   \int d\omega^Y \sum_{\alpha^Y}
   v(\omega^Y,\alpha^Y)|\omega^Y,\alpha^Y\rangle
\end{eqnarray*}
one can obtain the expression for the dissipation function from
\eq{12}:
\begin{equation}
   D(t,t_1) \equiv D_s(t-t_1) =
   \frac
   {\langle\tilde y z_0|
   e^{-i(H^{YZ}_0 + W_{YZ})(t-t_1)}
   |\tilde y z_0 \rangle}
   {\langle\tilde y z_0|
   e^{-iH^{YZ}_0(t-t_1)}
   |\tilde y z_0 \rangle}
   \label{eq24}
\end{equation}
and the expression for the function $\Delta_s(\epsilon)$ from \eq{16}:
\begin{equation}
   \Delta_s(\epsilon) =
   \frac{1}{\pi}{\rm Re}\int_0^\infty D_s(\tau) e^{-i\epsilon\tau}
   d\tau =
   \frac{1}{2\pi}\int_{-\infty}^{+\infty} D_s(\tau)
   e^{-i\epsilon\tau} d\tau.
   \label{eq25}
\end{equation}
The subscript $s$ is an abbreviation of ``scattering''. As one can
see from \eq{23}, $M(\omega)$ in \eq{15} depends on $\omega^Y$ only.
Thus, we can rewrite \eq{15} as
\begin{equation}
   \Gamma =
   2\pi \int d\omega^Y M(\omega^Y)\Delta_s(\omega^Y - \omega^Y_f)
   \label{eq26}
\end{equation}
where $\omega^Y_f = \omega^Y_0 + \omega^X_0 - \omega^X_1 = \omega^Y_0
+ \omega_{01}$ is the expected value of the final energy of particle
$Y$ in accordance with the energy conservation law. The formulae
(\ref{eq26}), (\ref{eq25}), (\ref{eq24}) show the perturbed value of
decay rate for considered problem and coincide with the formulae
(26), (27) and (18) of \cite{PANOV96F} for the same case.  Further
analysis of this formulae can be found in \cite{PANOV96F}.

\section{Decay onto unstable level}
\label{UNSTABLE}

Let's consider three-level system $X$ which makes a cascade
transition from the initial state $|x_0\rangle$ to the state
$|x_1\rangle$ and then to the state $|x_2\rangle$ with emission of
two particles $Y$ and $Z$, respectively. So, what is the influence of
instability of state $|x_1\rangle$ on decay constant of the state
$|x_0\rangle$? Again, it is a particular case of general situation
described in Section \ref{EVALUATION}. We consider particles $Y$ and
$Z$ as separate systems which are in the initial vacuum states
$|y_0\rangle$ and $|z_0\rangle$ at the moment $t=0$ and then they
transit to continuum $\{|\omega^Y,\alpha^Y\rangle\}$ and
$\{|\omega^Z,\alpha^Z\rangle\}$, respectively. The transition
$|x_0\rangle\to|x_1\rangle$ is caused by interaction $V =
V_{XY}\otimes I_Z$, and transition $|x_1\rangle\to |x_2\rangle$ is
caused by interaction $W = W_{XZ}\otimes I_Y$, where
\begin{eqnarray}
   V_{XY} &=& \int d\omega^Y \sum_{\alpha^Y}
   |x_1\rangle|\omega^Y,\alpha^Y\rangle\langle x_0|\langle y_0|
   v(\omega^Y,\alpha^Y) + {\rm E.C.}
   \label{eq27}\\
   W_{XZ} &=& \int d\omega^Z \sum_{\alpha^Z}
   |x_2\rangle|\omega^Z,\alpha^Z\rangle\langle x_1|\langle z_0|
   w(\omega^Z,\alpha^Z) + {\rm E.C.}
   \label{eq28}
\end{eqnarray}
It is seen from eqs.~(\ref{eq27},\ref{eq28}) that $V$ causes
transition $|x_0\rangle\to |x_1\rangle$ only and $W$ causes
transition $|x_1\rangle\to |x_2\rangle$ only. This process is
characterized by relations
\begin{displaymath}
   \{|\xi\rangle\}  =
   \{|x_1\rangle|\omega^Y,\alpha^Y\rangle|z_0\rangle\}; \quad
   \{|\eta\rangle\} =
   \{|x_2\rangle|\omega^Y,\alpha^Y\rangle|\omega^Z,\alpha^Z\rangle\}
   \label{eq29}
\end{displaymath}
and by a number of relations, coinciding with
eqs.~(\ref{eq17},
\ref{eq18},
\ref{eq19},
\ref{eq20},
\ref{eq23},
\ref{eq23p}).

It is not difficult to show that the dissipation function \eq{12}
now has the form
\begin{equation}
   D(t,t_1) \equiv D_u(t-t_1) =
   \langle x_1 z_0| e^{-i(H^{XZ}_0 + W_{XZ})(t-t_1)}|x_1 z_0\rangle
   e^{i{\cal E}^{XZ}_0(t-t_1)},
   \label{eq30}
\end{equation}
where $H^{XZ}_0 = H^X_0 \otimes I_Z + I_X \otimes H^Z_0$ and ${\cal
E}^{XZ}_0 = \omega^X_1 + \omega^Z_0$. The subscript $u$ is an
abbreviation of ``unstable''. We see from \eq{30} that the
dissipation function now has a simple physical sense.  This is
nothing more than a no-decay amplitude of the sate $|x_1\rangle$
in relation to decay under the influence of interaction $W$.  Thus, we
can use the approximation
\begin{equation}
   D_u(\tau) = e^{-\lambda\tau},
   \label{eq31}
\end{equation}
where $\lambda$ is a complex decay constant of level $|x_1\rangle$.
Let $\lambda = \lambda_r - i\lambda_i$, where $\lambda_r$ and
$\lambda_i$ are real numbers. Then we obtain from \eq{31} and \eq{16}
\begin{equation}
   \Delta_u(\epsilon) = \frac{1}{\pi}
   \frac{\lambda_r}{\lambda_r^2 +(\epsilon-\lambda_i)^2}
   \label{eq32}
\end{equation}
and we obtain from \eq{15} and \eq{32}
\begin{equation}
   \Gamma = 2\pi\int d\omega^Y M(\omega^Y) \frac{1}{\pi}
   \frac{\lambda_r}{\lambda_r^2 +
   \left[\omega^Y -\left(\omega ^Y_f + \lambda_i\right)\right]^2}.
   \label{eq33}
\end{equation}
with the same notations as in \eq{26}. For the special case
$\lambda_i = 0$, $\lambda_r \ll \omega^Y_f$, $M(\omega^Y) = {\rm
const}$ for $\omega^Y > 0$ we obtain the formula, similar to eq.~(20)
in \cite{ZEN_MENSKY99A}.

\section{Rabi transition from final state of decay}
\label{RABI}

Now we analyze the last particular case of general problem of Section
\ref{EVALUATION}. The situation is similar to that described in
Section \ref{UNSTABLE}, but the instability of the state
$|x_1\rangle$ is caused by a forced resonance Rabi transition to
another state $|x_2\rangle$.  We describe Rabi transition
semiclassically by the time-dependent interaction
\begin{displaymath}
   W_X(t) = \Omega (|x_1\rangle\langle x_2| + |x_2\rangle\langle
   x_1|) \cos\omega_{21}t
   \label{eq34}
\end{displaymath}
where $\omega_{21}$ = $\omega^X_2 - \omega^X_1$ and $\Omega$ is the
Rabi frequency. Spontaneous transition $|x_0\rangle \to |x_1\rangle$
is described in same manner as in the previous sections. We have
\begin{eqnarray*}
   S &=& X \otimes Y\\
   H &=& H^X_0 \otimes I_Y + I_X \otimes H^Y_0 + V_{XY} +
   W_X(t) \otimes I_Y\\
   |\Psi_0\rangle &=& |x_0 y_0\rangle; \quad
   {\cal E}_0 = \omega^X_0 + \omega^Y_0\\
   \{|\xi\rangle\} &=& \{|x_1\rangle|\omega^Y,\alpha^Y\rangle\}; \quad
   \{|\eta\rangle\} = \{|x_2\rangle|\omega^Y,\alpha^Y\rangle\} \\
   \omega &=& \omega^X_1 + \omega^Y.
\end{eqnarray*}
Based on \eq{12}, it's not difficult to show that the dissipation
function is
\begin{equation}
   D(t,t_1) = \langle x_1|x(t)\rangle e^{i\omega^X_1(t-t_1)}
   \label{eq35}
\end{equation}
where $|x(t)\rangle$ is the solution of equation
\begin{equation}
   |\dot x(t)\rangle =
   -i\left[H^X_0 + W_X(t)\right]|x(t)\rangle, \quad
   |x(t_1)\rangle = |x_1\rangle.
   \label{eq36}
\end{equation}
Using the rotating wave approximation we find from \eq{36}
\begin{equation}
   \langle x_1| x(t)\rangle = \cos\frac{\Omega}{2}(t-t_1)
   e^{-i\omega^X_1(t-t_1)}.
   \label{eq37}
\end{equation}
Substituting the scalar product from \eq{37} in \eq{35} we obtain
\begin{equation}
   D(t,t_1) \equiv D_R(t-t_1) = \cos\frac{\Omega}{2}(t-t_1).
   \label{eq38}
\end{equation}
The subscript $R$ is an abbreviation of ``Rabi''. Using \eq{38} and
\eq{16} we find
\begin{displaymath}
   \Delta_R(\epsilon) = \frac{1}{2}
   \left[ \delta\left(\epsilon - \frac{\Omega}{2}\right) +
   \delta\left(\epsilon + \frac{\Omega}{2}\right) \right]
   \label{eq39}
\end{displaymath}
and then we obtain from \eq{15}
\begin{equation}
   \Gamma = \pi \left[ M\left(\omega^Y_f - \frac{\Omega}{2}\right) +
    M\left(\omega^Y_f + \frac{\Omega}{2}\right)\right].
    \label{eq40}
\end{equation}
The resultant \eq{40} coincides with conclusion of
\cite{ZEN_PASCAZIO97} (see \cite{ZEN_PASCAZIO97}, eq.~(2.31)). But
the method by which this conclusion has been obtained in
\cite{ZEN_PASCAZIO97} significantly differs from our method.  The
forced transition from the level $|x_1\rangle$ (in the terms of
present paper) to the level $|x_2\rangle$ was described by means of
full quantum method, using quantized electromagnetic field instead of
classical potential, rather than semiclassically.  Instead of value
$\Omega/2$ in our \eq{40}, the quantity $B$ arises in eq~(2.31)
\cite{ZEN_PASCAZIO97}:
\begin{displaymath}
   B = |\Phi_0|\sqrt{N_0}
   \label{eq41}
\end{displaymath}
where $\Phi_0$ is the transition matrix element and $N_0$ is the
number of field quanta in resonance with $|x_1\rangle\to|x_2\rangle$
transition. But it is not difficult to show that value $B$ is
precisely the half of Rabi frequency. Thus, full quantum and
semiclassical methods produce the same results. One can note that
full quantum description of Rabi transition can be treated in the
frame of general formalisms of Section \ref{EVALUATION} as well.

\section{Discussion}
\label{DISCUSSION}

It is easily seen from the formulae (\ref{eq26}), (\ref{eq33}),
(\ref{eq40}) that, in the formal limit of very fast dissipation of
the final state of decay caused by interaction with environment, the
spontaneous exponential decay is frozen. Really, the function
$M(\omega)$ has only finite width. Very fast dissipation of the final
state means that function $\Delta(\epsilon)$ becomes very wide.  But
$\int\Delta(\epsilon)d\epsilon = 1$ in all cases. Thus, the integral
in the right hand side of eqs.~(\ref{eq26}), (\ref{eq33}),
(\ref{eq40}) tends to zero as the width of function
$\Delta(\epsilon)$ tends to infinity, consequently $\Gamma\to 0$.
This is an expression of Zeno paradox in dynamical consideration
(without using projection postulate).  But dissipation of final
states of decay does not necessarily cause decrease of decay
constant.  When dissipation is not very strong, the behavior of decay
constant may be rather complex, its behavior depends on fine features
of the function $M(\omega)$.  For example, if we consider realistic
case $\Omega \ll \omega_{01}$ for Rabi transition from the final
state of usual electromagnetic transition, the relation of disturbed
to no-disturbed decay constant is
$$
   \frac{\Gamma}{\Gamma_0} = 1 +
   \frac{3}{4}\frac{\Omega^2}{\omega_{01}^2}
$$
i.~e.\ $\Gamma > \Gamma_0$.  And only in the case of very fast
dissipation of final state, decay constant starts to decrease.

Let's note that the decay constant perturbation by Rabi transition
from final decay state (Section \ref{RABI}) seems to be not an usual
QZE, because there are no irreversible changes in the environment
following such transition, consequently there is no event of
measurement.  This is clearly seen from the semiclassical picture of
this phenomenon. But such phenomenon is closely related to QZE and we
can consider it as a Zeno-like effect. Consequently our main formula
(\ref{eq15}) describes not only Zeno effect itself, but a wide class
of Zeno-like effects. There are also some other phenomena which can
be described in the frame of general theory of Section
\ref{EVALUATION}, but have been excluded from our consideration:
spontaneous oscillation in the final states of decay or combinations
of different mechanisms considered above are examples of such
phenomena.

\vskip 0.5cm
\centerline{\bf ACKNOWLEDGMENTS}

The author acknowledges the fruitful discussions with M.~B.~Mensky.
The work was supported in part by the Russian Foundation of Basic
Research, grant 98-01-00161.

\end{document}